\documentclass[11pt]{article}
\usepackage{geometry}                % See geometry.pdf to learn the layout options. There are lots.
\usepackage{graphicx}
\usepackage{amssymb}
\usepackage{amsmath}
\usepackage{amsfonts}
\usepackage{graphicx}
\usepackage{epstopdf}
\usepackage{hyperref}
\usepackage{color}

\usepackage{eso-pic}

\usepackage{cite}
\newcommand{\E}{\mathbb{E}}

\newcommand{\be}{\begin{equation}}
\newcommand{\ee}{\end{equation}}
\newcommand{\bea}{\begin{eqnarray}}
\newcommand{\eea}{\end{eqnarray}}
\newcommand{\beas}{\begin{eqnarray*}}
\newcommand{\eeas}{\end{eqnarray*}}

\newcommand{\e}{\mathrm{e}}

\begin{document}

\title{Stability analysis of financial contagion\\ due to overlapping portfolios}
\author{Fabio Caccioli$^{1}$, Munik Shrestha$^{1,2}$, Cristopher Moore$^{1,2}$, and J. Doyne Farmer$^{3,1}$
~~~~~\\
{\em 1 - Santa Fe Institute, 1399 Hyde Park road, Santa Fe, NM 87501, USA}\\
{\em 2 -  University of New Mexico, Albuquerque, NM 87131 }\\ 
{\em 3 - Institute of New Economic Thinking and Mathematical Institute,}\\
{\it 24-29 St. Giles, University of Oxford, Oxford OX1 3LB, UK }\\ 
}

\maketitle
\begin{abstract}
Common asset holdings are widely believed to have been the primary vector of contagion in the recent financial crisis.  We develop a network approach to the amplification of financial contagion due to the combination of overlapping portfolios and leverage, and we show how it can be understood in terms of a generalized branching process.   By studying a stylized model we estimate the circumstances under which systemic instabilities are likely to occur as a function of parameters such as leverage, market crowding, diversification, and market impact.  Although diversification may be good for individual institutions, it can create dangerous systemic effects, and as a result financial contagion gets worse with too much diversification.  Under our model there is a critical threshold for leverage; below it financial networks are always stable, and above it the unstable region grows as leverage increases.   The financial system exhibits ``robust yet fragile'' behavior, with regions of the parameter space where contagion is rare but catastrophic whenever it occurs.  Our model and methods of analysis can be calibrated to real data and provide simple yet powerful tools for macroprudential stress testing.

\end{abstract}

\tableofcontents

\section{Introduction}
The 2007--2009 financial crisis highlighted the complex interconnections between financial institutions and made it clear that we need a better understanding of how financial contagion propagates and the circumstances under which it is amplified\cite{Gai10,Haldane11,Allen12,Ibragimov11,Gai11,May10,Amini10,Georg10, Caccioli12}.  Financial contagion comes through different channels, including {\em (i)} counterparty risk, {\em (ii)} roll-over risk, and {\em (iii)} common asset holdings, i.e. {\it overlapping portfolios}.  Of these the first two have so far received the most attention, even though the primary problem is believed by many to have been due to the third.  Our goal in this paper is to remedy this by gaining a better understanding of the problem of overlapping portfolios.  To do this we develop a method of computing the stability of financial networks under contagion due to overlapping portfolios.  To understand the factors that determine network stability, we develop and study a stylized model, and suggest how it can be extended to be more realistic.  This model can be regarded as a multiple asset extension of the single asset model developed in reference \cite{Caccioli12b}\footnote{
Reference \cite{Caccioli12b} considered the properties of leveraged single asset portfolios.  It was shown that under deleveraging market impact can cause bankruptcy if leverage is too large.}.

Inter-institutional lending drives the problem of counterparty and roll-over risk.  Counterparty risk occurs when a bankrupt institution is unable to pay its debts and consequently causes other institutions to fail \cite{Staum11}. Roll-over risk occurs when financial institutions depend on short term lending for liquidity and their creditors stop lending because they fail or are under stress, so that they are no longer able to borrow and consequently fail or become under stress \cite{Gai11}.  These have now been extensively studied and we are rapidly developing better insight into the circumstances where interbank lending causes problems (see for instance \cite{Gai10,May10,Staum11}).

Financial contagion due to overlapping portfolios is driven by common asset holdings \cite{May10,Beale11}.   In the event that an asset price fluctuation causes an institution to fail, the resulting ``fire sale'' of assets by that institution further depresses prices, which in turn may cause other institutions to fail, causing a spiral of selling and further asset price decreases.  This also induces correlations between different assets that further exacerbate the problem \cite{Cont12}.  

The problem of overlapping portfolios is very general.  It occurs even without inter-institutional lending, and applies to any institutions that manage money.   Although this can occur even without leverage\footnote{
In this paper we assume that institutions sell assets only when they become bankrupt, in which case the problem of financial contagion occurs only when leverage is used.  In general asset sales may be triggered by losses that are less severe, for example if investment funds are forced to liquidate even when they are solvent, as occurred during the stat-arb meltdown in 2007.},
the use of leverage makes it particularly acute.  We are particularly interested in the banking system, where it is not uncommon for investments to be leveraged by a factor of 30 or more, but our analysis applies equally well to hedge funds or any other financial institutions that make leveraged investments.  For convenience we will use the word 
\emph{bank} to refer to institutions in general, 
but the reader should bear in mind that our model applies equally well to any leveraged financial institution.

The problem of overlapping portfolios has previously been considered in references \cite{Gai10, May10, Nier08,Arinaminpathy12}\footnote{After completing these results we received reference \cite{Huang12}, whose independent results are complementary to ours.}. In these papers, however, liquidation effects were considered on top of counterparty or roll-over risk.
Here we are interested in the situation in which shocks can propagate between different financial institutions through a pattern of local portfolio overlaps (e.g.\ bank $i$ has assets in common with bank $j$, that has other assets in common with bank $k$, etc.).  The model is simple: we assume that banks 
own a portfolio of assets, 
that when a bank goes bankrupt due to a loss in the value of its portfolio it sells its assets, and that this in turn causes these assets to be devalued according to 
a simple market impact function relating the size of the sale to the change in price.

The model we consider is purely mechanistic, i.e.\ we do not attempt to describe decision-making processes by banks. The underlying assumption is that, during the development of a crisis, banks do not have time to deleverage or rebalance their portfolios before failing.  Thus we consider portfolios fixed until default occurs, and assume that they are fully liquidated when it occurs.  We then perform a macroprudential stress test by applying localized shocks affecting either a single bank or a single asset.  After the initial shock is applied we test to see whether it causes any bank failures; if so we iterate the process as needed until either there are no more failures or all banks have failed.  The only trades during the course of the dynamics are fire sales of the assets of insolvent banks. 

Our focus in this paper is therefore in understanding the specific role of market impact and portfolio overlaps as a contagion mechanism between leveraged financial institutions. 
To this end, we consider a network of banks and assets, 
and we test how the average level of diversification in bank portfolios, the 
ratio of the number of banks to the number of assets (crowding), and the leverage attained by banks impact the stability of the system with respect to an initial shock affecting a single asset or bank.

The stability of the system will be measured in terms of the probability of observing a global cascade of failures, with a smaller probability being associated with a higher stability.  A global cascade of failures, in this context, refers to the failure of a significant fraction of the banks: 
that is, a non-zero fraction in the limit of infinite network size. 
By mapping our model onto a generalized branching process, 
we show analytically that there is a region in parameter space where global cascades of failures occur.  
One advantage of this mechanistic approach is that it can in principle be calibrated against real data and used to perform stress tests on real financial systems.

We find that, as the diversification of the banks' portfolios increases, the system undergoes two phase transitions, with a region in between where global cascades occur.  
Below the first transition, banks are not interconnected enough for shocks to propagate in the network.  Above the second transition, banks are robust to devaluations in a few of their assets.  In between these two transitions, banks are both vulnerable to shocks in their asset prices, and interconnected enough for these shocks to spread.  
We also find that more leverage increases the overall instability of the network and that the system exhibits a ``robust yet fragile''  behavior, with regions of parameter space where contagion is rare but the whole system is brought down whenever it occurs.

The paper is organized as follows.  In the next section we introduce the model. In Section~\ref{theory} we map the model into a generalized branching process and present the analytical approach that allows us to identify the region of phase space where global cascades occur. In Section~\ref{simulations} we report results from numerical simulations exploring how stability of banking systems depends on parameters and network properties.  In section 5 we compare the results of numerical simulations to those of stability analysis, and we present our conclusions in the last section.

\section{The model}\label{model}

\subsection{Banks, assets, and cascades of bankruptcies}

\begin{figure}[h]
\begin{center}
\includegraphics[width=4in]{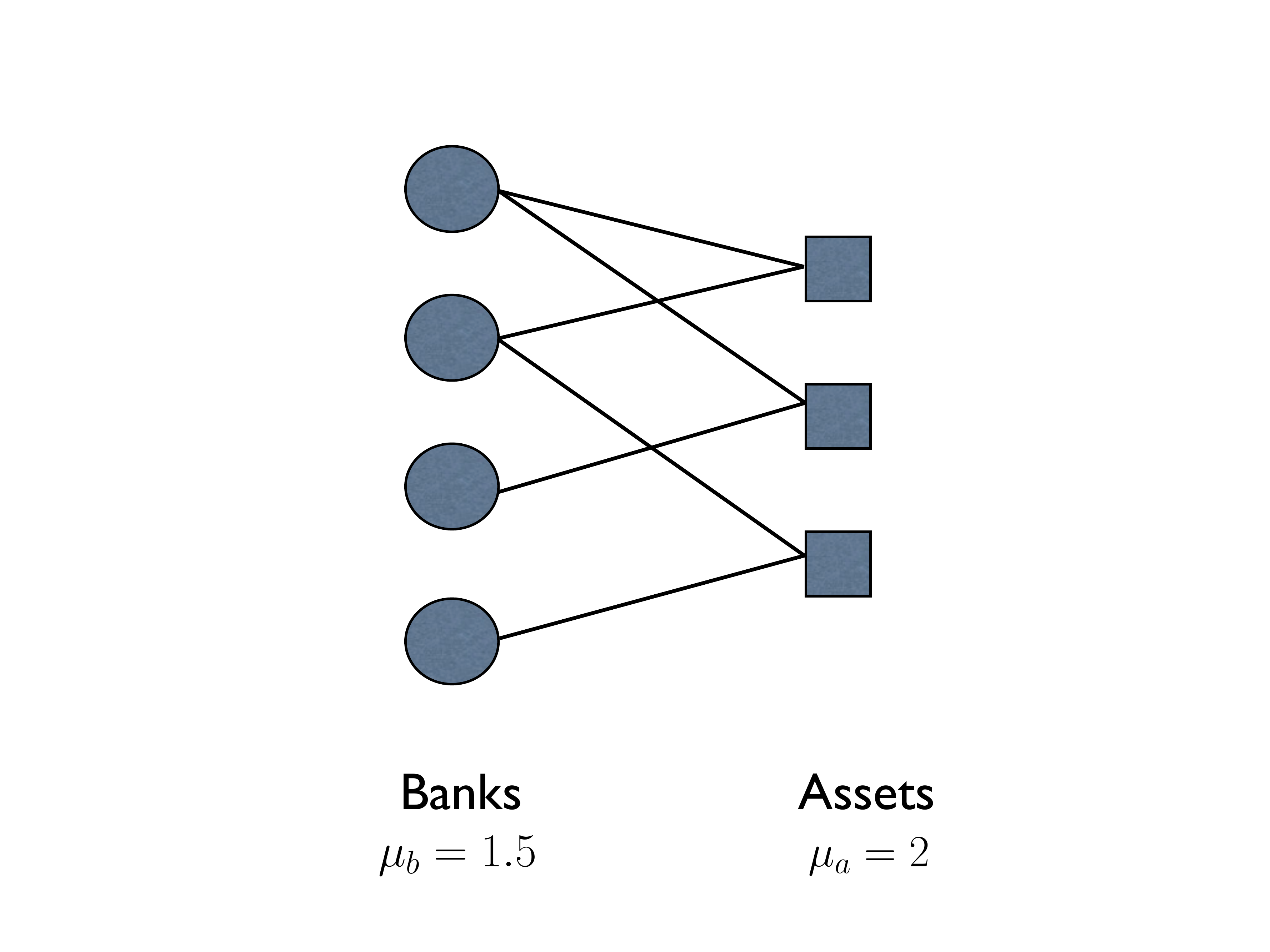}
\end{center}
\caption{\footnotesize{Graphical representation of the bipartite network of banks and assets. Banks are denoted by circles, assets by squares. Links connect banks to the assets they have in their portfolios. In this example $N=4$, $M=3$, the average banks' degree is $\mu_b=1.5$ and the average assets' degree is $\mu_a=2$.}}
\label{fig0}
\end{figure}

We consider a representation of a financial system given in terms of a network of $N$ banks and $M$ assets. 
Whenever a bank invests in an asset, we draw a link in the network connecting that bank to that asset.  
The resulting network is bipartite (see Figure~\ref{fig0}), meaning that there are two groups of nodes (banks and assets) and that there are links only between these two groups.  
The number of assets in the portfolio of bank $i$, i.e.\ the number of links of the corresponding node, is its degree $k_i$.  The {\it average diversification}, i.e.\ the average degree of banks in the network, is then
\be
\mu_b=\frac{1}{N} \sum_{i=1}^N k_i \, ,
\ee
where the sum runs over all $N$ banks.  
Conversely, 
the number of banks that hold asset $j$ in their portfolio is its degree $\ell_j$, and the average degree of the assets is
\be
\mu_a=\frac{1}{M} \sum_{j=1}^M \ell_j \, .
\ee 
Since each link connects a bank to an asset, the total degree of the banks must equal the total degree of the assets, so 
\be
\mu_b N=\mu_a M \, . 
\ee
Although a complete description of the network's topology would require more information, a rough characterization can be given in terms of two parameters, $\mu_b$ and $n=N/M$.  
The {\it crowding parameter} $n$ is a measure of the density of institutions choosing their investments from the same pool of assets.

Each solvent bank $i$ holds a portfolio $\{Q_{i,1},\ldots,Q_{i,M}\}$.  Its value at time $t$ is 
\[
A_i^t=\sum_{j=1}^M Q_{ij} p_j^t \, , 
\]
where $Q_{ij}$ is the number of shares of asset $j$ held by bank $i$ and $p_j^t$ the price of asset $j$ at time $t$.
In our dynamics a bank holds on its portfolio as long as it is solvent, so $Q_{ij}$ is independent of time. Notice that, given that bank $i$ invests in $k_i$ assets, only $k_i$ of the $M$ portfolio weights $Q_{ij}$ will be non-zero for bank $i$.

Each solvent bank also holds cash $C_i$, and we denote by $L_i$ its total liabilities; neither of these quantities depend on time. If $A_i^0$ is the initial value of bank $i$'s portfolio, its initial equity (or capital) is therefore $E_i^0=A_i^0+C_i-L_i$.
The leverage of a bank is the ratio between the amount of risky assets on its balance sheet and its equity. Assuming no risk associated with cash holdings, the initial leverage of bank $i$ is $\lambda_i=A_i^0/E_i^0$.

The condition for bank $i$ to be solvent at time $t$ is
\be
\sum_{j=1}^M Q_{ij} p_j^t+C_i\ge L_i \, .
\ee
Given that $E_i^0=A_i^0+C_i-L_i$, the above condition can be expressed as
\be\label{solvency}
A_i^0-\sum_{j=1}^M Q_{ij} p_j^t \le E_i^0 \, . 
\ee
The left hand side represents the loss with respect to the initial investment.  If such a loss happens to be greater than the initial capital of the bank, the bank is out of business.  

Note that leverage is a necessary condition for  banks to fail.  A bank investing only its own capital always satisfies condition~\eqref{solvency}, since its maximal loss is equal to its equity.  We can write \eqref{solvency} as a condition on the leverage, 
\be
\lambda_i \le \frac{\sum_{j=1}^M Q_{ij} p_j^t}{E_i^0}+1 \, .
\ee
Even in the worst case scenario where $p_j^t=0$ for all assets, this condition can be violated only if $\lambda_i > 1$, i.e.\ if the bank is leveraged.

Whenever a bank does not satisfy the solvency condition \eqref{solvency}, we assume its portfolio undergoes a fire sale, i.e.\ all its assets are immediately liquidated. The fire sale causes the price of the assets in the bank's portfolio to drop. 
If   $x_j$ is the fraction of asset $j$ that has been liquidated, the price is updated as
\be
p_j\to p_jf_j(x_j)
\ee

We are interested in the response of the system to an initial shock. 
We consider two kinds of initial shocks:
\begin{itemize}
\item {\bf Presence of a toxic asset.} We select a random asset $j$ and devalue it at time $0$.  
\item {\bf Initial failure of a bank.} We select a random bank $i$ and cause it to go bankrupt. 
\end{itemize}
In each case we follow the chain of events caused by the initial shock.  
The dynamics we consider is very simple: after shocking the system at time $t=0$, at each time step $t=1,2,\ldots$ the solvency condition~\eqref{solvency} is checked for each bank, the portfolios of newly insolvent (bankrupted) banks are liquidated, and new prices are computed for each asset.  
The dynamics stops when no new bankruptcies occur between two consecutive time steps.  
This can be expressed with the following algorithm:
\begin{enumerate}
\item introduce the initial shock in the system;
\item liquidate the portfolio of insolvent banks;
\item recompute prices of assets;
\item if new banks are insolvent go to step 2, otherwise end.
\end{enumerate}
\noindent
Note that we don't allow for new banks to enter the system, so that once a bank has gone bankrupt it remains in this state for the rest of the process. 

In the limit of large systems, when $N,M \to \infty$ while the parameters $\mu_b$ and $n=N/M$ remain finite, the initial shock we consider only affects an infinitesimal (of order $\mathcal{O}(1/N)$) fraction of the banking system. We are interested in understanding if and when such infinitesimal shocks can trigger global cascades of failures. A {\it global cascade} of failures is defined as a cascade affecting a finite fraction of banks in the infinite system. 
In the following we will measure the probability and the average extent of contagion.
We define the {\it probability of contagion} as  the probability that a global cascade of failures occurs, and the {\it average extent of contagion} as the average size of a global cascade.

\section{Stability analysis}\label{theory}

In this section we develop a theoretical approach that allows us to compute a bound on stability, which as we will show is a good estimate, for when cascading bank failures are likely to occur.  We show how this can be applied to understand the stability of specific banking networks (i.e.\ a given set of banks and their balance sheets),  and we also show how it can be used to understand how stability depends on the parameters of the network, 
such as diversification, crowding, and leverage.

Let us start by discussing what happens if there is an external shock that causes a particular bank to go bankrupt. Through the combination of leverage and impact, this failure can trigger the failure of other banks investing in the same assets.  If the parameters of the system amplify shocks, this can generate a cascading failure that propagates through the system.   One of our main points is that, while the likelihood of the first bank failure depends on the nature of the shocks, whether or not this propagates depends on whether the financial system is stable, which in turn depends on parameters such as the leverage, market impact and network structure.  We begin with a general discussion of branching processes.  We then discuss how it can be applied to understand a given banking network, and make some specific assumptions that allow us to demonstrate how the stability of the banking system depends on parameters.  

\subsection{The Galton-Watson process}

The scenario of cascading failures for banks closely resembles the branching process introduced  by Galton and Watson to study the survival probability of family names over generations \cite{Galton75}.  This process is formulated in terms of a progenitor that gives rise to $x$ children, where $x$ is a non-negative integer drawn from a probability distribution $g(x)$.  Each of the children, in turn, independently generates a number of offspring distributed according to $g(x)$, and the same process is repeated at each generation. The question is whether such a process is doomed to extinction or not, i.e.\ if the population drops to zero after a finite number of generations, so that the total number of descendants is finite.  
A fundamental result in the theory of branching processes states that such a process goes extinct with probability one if $\E[x] < 1$, where $\E[x]$ is the expected number of offspring per individual.  

For our purposes it is essential to consider a generalized Galton-Watson process with individuals of different types $i\in{1,2,\ldots,q}$.  The key quantities are then, for each pair of types $i, j$, the expected number of offspring of type $i$ produced by an individual of type $j$.  We denote these as a $q \times q$ matrix $\mathcal{N}_{ij}$.  
The condition for extinction is then that the largest eigenvalue $\xi_1$ of $\mathcal{N}$ is smaller than one~\cite{Mode71}.  
Conversely, if this eigenvalue is greater than one, then with positive probability this process lasts forever, producing an infinite number of offspring.  We say that the branching process is \emph{subcritical} or \emph{supercritical} if this eigenvalue is less than or greater than one, respectively.  

In the context of our model, we are interested in computing the expected number of banks that go bankrupt because of the previous failure of another bank. Consider for example the case in which a random bank $i$ receives a shock at time $t=0$ that causes it to become bankrupt. This bank is equivalent to the progenitor of the Galton-Watson process, and banks whose bankruptcy is triggered by that of $i$ are equivalent to its offspring.  In the language of branching processes, banks failing at time $t$ correspond to  individuals in the $t$-th generation.  We are interested in understanding when there is a non-zero chance that financial contagion keeps on spreading over time, which is equivalent to asking whether shocks will be amplified 
rather than dying out.  If the branching process is supercritical, then this initial shock results in a global cascade with non-zero probability, affecting a non-zero fraction of all the banks in the limit of infinite system size.  

Note that in our model we have in principle banks with different properties (degree, leverage, size\ldots) that can be considered as individuals of different types in the generalized Galton-Watson process.  Thus $\mathcal{N}_{hk}$ is the expected number of banks of type $h$ that fail because of the failure of a bank of type $k$.  
There is obviously considerable flexibility in how we classify banks into types, which at its most fine-grained extreme allows the ``types'' to correspond to individual banks.

It is important to stress at this point that the process here considered is more complex than the usual generalized Galton-Watson process. In particular, in our case, the ultimate fate of bank $i$ depends not only on its properties, but also on those of all the other banks whose portfolio overlaps with $i$. This happens because the price drop that follows the fire sale liquidation of an asset depends on the fraction of total shares of that asset being liquidated, which changes from bank to bank. 

A second important difference is that the Galton-Watson process occurs on a tree, so that individuals of a given generation are independent.  For banks the failure process is not necessarily a tree, but is rather a more general graph which may have loops.  To see this, consider a simple example of three banks $i$, $j$, and $k$ with one asset in common. Let us suppose that $i$ is robust with respect to the failure of $j$ by itself, but not with respect to the failure of both $j$ and $k$ together. Now, if the failure of $j$ is enough to trigger that of $k$, then $i$ is effectively vulnerable to the failure of $j$. Such situations, which occur whenever assets have degree higher than $2$, are neglected under the analytical calculation that we perform here. Therefore, our analytical treatment gives a sufficient but not a necessary condition for global cascades to occur, and gives only an upper bound on the stability of the banking system.   We will see, however, that it is nonetheless a good approximation, in rough agreement with the results of numerical simulations. 

It is in principle possible to improve this approximation to account for the non-linearities induced by loops in the branching process by considering multiple time-step dynamics.  This method is commonly used in dynamical system theory:  the $t$-th iteration of the dynamics converts cycles of length $t$ into fixed points.  For instance, to properly treat triplets one can compute a two-step matrix that counts the average number of banks of type $i$ whose failure is triggered by the bankruptcy of a bank of type $j$ within two time steps of the dynamics.  Comparing to the example given above, if an initial shock causes $j$ to fail, $k$ will fail after one iteration, and since both $j$ and $k$ have now failed, $i$ will fail in the second time step.  
While our one-step approximation is already quite accurate, this approach 
provides a path for systematically improving the degree of approximation, which deserves further investigation.

\subsection{Stability of a given system}

If we have complete information about the banking system, i.e.\ if we know the portfolios $Q_{ba}$ of all the banks and, in addition, the market impact function for their assets, then we can 
describe the stability of the system through a matrix $\mathcal{B}$, where $\mathcal{B}_{ij}$ is the probability that bank $i$ will fail under the failure of bank $j$.   Bank $i$ becomes insolvent when bank $j$ fails if and only if the market impact due to the sale of their overlapping assets causes a loss to bank $i$ that exceeds its equity $E_i$.  
As described above, we focus for now on the direct effect on bank $i$ of the failure of bank $j$.  
Since we assume that bank $j$'s entire portfolio $\{Q_{ja}\}$ is liquidated, the new price for asset $a$ is $p_a(1 -  f_a(Q_{ja}))$.  
Using the shorthand $\textrm{Prob}(x)$ to indicate the probability that condition $x$ is satisfied, the stability matrix $\mathcal{B}_{ij}$ is defined as
\be
\mathcal{B}_{ij} = \textrm{Prob} 
\left[ \sum_{a=1}^M  Q_{ia} p_a  \left( 1 - f_a \left(Q_{ja} \right) \right) - E_i >0\right] \, .
\ee
In order to understand whether a cascade of failures will spread, we compute $\mathcal{B}_{ij}$ in the case where the assets shared by banks $i$ and $j$ have not yet been devalued, and still have their initial prices.  That is, we focus on the ``boundary'' of the cascade, with failures and devaluations spreading outward through the network through banks and assets that have not yet been touched by the crisis.  In that case, since the dynamics themselves are deterministic, $\mathcal{B}_{ij}$ depends only on the initial structure of the banks' portfolios, and in particular on the network structure.  
The stability of the banking system can then be estimated by simply computing the largest eigenvalue $\xi_1$ of $\mathcal{B}$ and determining whether $\xi_1$ is greater than or less than one.

Note that, rather than using the simplifying approximation that the market impact function is deterministic, 
one could more realistically use a stochastic market impact function as in~\cite{Caccioli12b}.  Similarly, imperfect knowledge about bank portfolios and equity can be coped with using probabilities to represent uncertainties in their values.  
In either case, we can still bound the stability of the network by computing $\mathcal{B}$'s largest eigenvalue.

\subsection{Simplifying assumptions}\label{assumptions}

The approach described above makes it possible to estimate the stability of the banking system when it is in a particular state, corresponding to a particular configuration of the balance sheets of each bank.  One of our main goals here, however, is to understand more generically how the stability of the banking system depends on its network properties.   To make a high-level characterization it is necessary to think in terms of ensembles of networks, and to understand how stability varies as properties of the ensemble are varied.  As a first step in this direction we will make some specific assumptions in order to simplify the problem and gain intuition.  While these assumptions are rather arbitrary, the basic method used here is easily generalized, as discussed later.

\begin{itemize}
\item {\bf Network topology:}  We will consider random networks with Poisson degree distributions for both banks and assets.  Specifically, for each possible bank-asset pair a link is drawn with probability $\mu_b/M$.  The resulting network is drawn from the bipartite Erd\H{o}s-Renyi ensemble of random networks with average degrees $\mu_b$ and $\mu_a=\mu_b N/M$ for the banks and assets respectively.
\item {\bf Structure of balance sheets:} We will assume all banks have the same amount of money $A_i^0=A^0$ available for investment, and that each bank uniformly splits its investment in the assets that are in its portfolio. The asset side of bank's balance sheets will be composed of $80\%$ assets and $20\%$ cash.  For bank $i$ each link thus corresponds to an investment of $0.8 A^0/k_i$, where $k_i$ is the number of assets in $i$'s portfolio. Unless otherwise stated, we assume for each bank an initial equity $E_i^0=E^0$ corresponding to $4\%$ of its total assets. This corresponds to all banks having initial leverage $\lambda=A^0/E^0=20$.
\item {\bf Market impact function:} We will assume that the market impact function has the form $f_j({x_j^t}) = \e^{-\alpha x_j^t}$, where $x_j^t$ is the fraction of asset $j$ liquidated up to time $t$.  The parameter $\alpha$ is chosen such that the price drops by $10\%$ when $10\%$ of the asset is liquidated, i.e.\ $\alpha=1.0536$. All prices are set to $p_j^0=1$ at time $0$. This choice corresponds to linear market impact for log-prices, as originally used to describe price dynamics in \cite{Bouchaud98,Farmer02}.  It should be noted that recent empirical and theoretical evidence indicates that market impact for large trades is a concave function of the number of traded shares, which under normal conditions impact is well approximated by a square-root function \cite{Bouchaud08b}.  By normal conditions we mean that execution is slow enough for the order book to replenish between successive trades.  Under extreme conditions, like those of a fire sale, market impact is expected to become less concave and even linear or super-linear \cite{Gatheral10}, which motivates our choice of functional form here.
\end{itemize}

Altering these assumptions does not change the qualitative behavior of the system.  In particular, our methods generalize easily to degree distributions other than Poisson, e.g.\ power laws, and also to multiple types of banks with different sizes, portfolio structures, and amounts of leverage, or multiple types of assets with different initial prices and market impact functions.

\subsection{Explicit calculation of the stability matrix} 

In order to understand how stability depends on network properties, we lump banks into equivalence classes according to their degree, equating their degree with their type in the generalized Galton-Watson process.  We define the following notation:
\begin{itemize}
\item
$N_h$ is the number of banks of degree $h$.
\item
$\mathcal{P}(h,k | a)$ is the probability that a given bank of degree $h$ and a given bank of degree $k$ share a given asset $a$, i.e., are both connected to $a$ in the network.
\item
$F(h | k, a)$ is the probability that a bank of degree $h$ fails given that it is connected to a failed bank of degree $k$ through asset $a$.
\end{itemize}
Under the assumption that we are in the limit where $M \to \infty$, $N \to \infty$ while $\mu_b$ and $n = N/M$ are finite, the network is sparse, and we can easily compute the probability $\mathcal{B}_{ij}$ by summing over each asset one at a time.  If $i$ has degree $h$ and $j$ has degree $k$, the probability that the failure of bank $j$ causes bank $i$ to fail can be written 
\be
\mathcal{B}_{ij} = \sum_a \mathcal{P}(h,k | a) \,F(h | k,a) \, . 
\ee
Summing over all banks of degree $h$, the expected number of failures of banks of degree $h$ caused by the failure of a bank of degree $k$, is 
\be\label{Nmatrix}
\mathcal{N}_{hk} = N_h \sum_{a=1}^M \mathcal{P} (h,k | a)F(h | k, a) \, .
\ee
This is the matrix defining the branching process, i.e.\ the expected number of offspring of type $h$ from an individual of type $k$. 

We can now compute each of the entires of $\mathcal{N}_{hk}$ in turn.  Since the degree distribution of our network ensemble is Poisson, the number of banks of degree $h$ is simply $N_h = N P_b(h)$ where 
\be
P_b(h)=\frac{\e^{-\mu_b} \mu_b^h}{h!}
\ee
is the probability that a bank has degree $h$.  
A given bank of degree $h$ is connected to a given asset $a$ with degree $\ell_a$ with probability $h\ell_a/(\mu_b N)$, where $\mu_b N$ is the total number of edges in the network. The probability that a failed bank of degree $k$ is also connected to the same asset $a$ is $h (k-1) \ell_a (\ell_a-1)/(\mu_b N)^2$, where the factor of $k-1$ comes from the fact that one of the $k$ edges of the failed bank is already connected to the asset that caused its failure. This gives
\be
\mathcal{P}(h,k | a)=\frac{h\ell_a(k-1)(\ell_a-1)}{\mu_b^2N} \, .
\ee

We now compute the probability $F(h | k, a)$ that a bank $i$ of degree $h$ fails due to failure of a bank $j$ of degree $k$ given that they share an asset $a$.   The shift in price when a fraction $x_a$ of an asset is sold is $(1 - f_a(x_a))$.  (Recall that the initial price is set to one for convenience).  Thus the condition for a bank of degree $k$ to fail because bank $j$ sells a fraction $x_a$ of asset $a$ is  
\be
\frac{A^0}{k}\left(1-f_a(x_a)\right) > E^0. 
\ee
If $\nu(a)$ denotes the set of banks investing in asset $a$, the fraction of $a$ that is liquidated when $j$ fails is
\be 
x_a = \frac{A^0/k}{\sum_{m\in\nu(a)} A^0/k_m}
= \frac{1/k}{\sum_{m\in\nu(a)} 1/k_m} 
= \frac{1/k}{1/h + 1/k + \sum_{m\in\nu'(a)} 1/k_m} \, ,
\ee
where $\nu'(a)$ denotes the set of banks, \emph{other than $i$ and $j$}, that invest in $a$ and $k_m$ the degree of bank $m$.  

To compute $F(h| k, a) $ we must add up the probability of failure for each possible configuration of banks that are compatible with the condition of choosing a specific pair of banks of degrees $h$ and $k$ that are connected through asset $a$.   If $a$ has degree $\ell_a$, there are $\ell_a-2$ remaining banks.  Letting $i$ index these banks, we must average over the possible configurations $\{m_1, \ldots, m_{\ell_a - 2}\}$.   Fortunately the degrees of the banks are independent.  The probability that bank $i$ has degree $m_i$ is the ratio of the number of edges for banks of degree $m$ to the total number of edges.  Since $N_m = N P_b(m)$,  the number of edges for banks of degree $m$ is $m N_m$, and the total number of edges in the network is $\mu_b N$.  Thus each bank has degree $m$ with probability $m N_m / (\mu_b N) = m P_b(m) / \mu_b$ and the probability of any given configuration of bank degrees  is 
\be
\prod_{i=1}^{\ell_a-2} \frac{m_i P_b(m_i)}{\mu_b}.
\ee
Combining equations (13 - 15) and summing over all the possible configurations $\{m_1, \ldots, m_{\ell_a - 2}\}$ gives
\be
F(h|k,a)=\\
\sum_{m_1 = 1}^\infty\cdots\sum_{m_{(\ell_a-2)} = 1}^\infty 
\prod_{i=1}^{\ell_a-2} \frac{m_i P_b(m_i)}{\mu_b}
\,\Theta\!\left[ \frac{A^0}{h}\left(1-f_a\left(\frac{1/k}{1/h + 1/k + \sum_{i=1}^{\ell_a-2} 1/m_i} \right) \right)-E^0 \right] \, ,
\ee
where $\Theta$ is the Heaviside step function, $\Theta(x)=1$ if $x > 0$ and zero otherwise.  

After summing over assets equation \eqref{Nmatrix} becomes

 \be\label{matrixm1}
\mathcal{N}_{hk}=\frac{\e^{-\mu_b} \mu_b^h}{h!}
  \frac{h  (k-1)}{\mu_b^2 n} \sum_{\ell} \frac{\e^{-\mu_a} \mu_a^\ell}{\ell!} \ell (\ell-1)F(h,k,\ell),
\ee
where we have used the fact that the number of assets with given degree $\ell$ is $MP_a(\ell)$ and explicitly introduced the Poisson degree distributions of banks and assets. 

The form of the matrix $\mathcal{N}$ confirms that the independent parameters of the model are $\mu_b$, $n$, $\lambda$ and $\alpha$. We can see in particular that, although
leverage has a similar effect on stability to the market impact constant $\alpha$, the two are not related through a simple relation that allows us to eliminate one of the two dependencies.  However, if we had used a market impact that was linear in the price, instead of the log-price, i.e.\ of the form $f_a(x) = \alpha' x$, then the stability would depend only on the product $\alpha' \lambda$ and not on the two parameters separately. 

For networks in which all the banks have the same degree $k$ we can compute the largest eigenvalue of $\mathcal{N}$ in closed form. In this case the matrix $\mathcal{N}$ reduces to the scalar quantity
\be
\mathcal{N}=\xi_1=(k-1)\mu_b n\frac{\Gamma(l^*-1,\mu_b n)}{\Gamma(l^*-1)},
\ee
where
\be
l^*=\frac{1}{\log\left(\frac{\lambda}{\lambda-k}\right)},
\ee
$\Gamma(x)$ is the gamma function, and $\Gamma(x,z)$ is the incomplete Gamma function. 

If we make the approximation $\frac{1}{k_b}\to\E\left[\frac{1}{k}\right]$ we can obtain a closed expression for $\mathcal{N}_{h,k}$.   However, given that this approximation is uncontrolled, we do not give an explicit form for the matrix elements, but rather compute them exactly via Montecarlo methods.

\section{Dependence on leverage and network properties}\label{simulations}

We now explore how the stability of the banking network depends on parameters.  We first show results based on numerical simulations and then compare them to results based on the stability matrix $\mathcal{N}$.  

In numerical simulations $N$ and $M$ are both finite, and global cascades can be defined as cascades for which the fraction of bankrupted banks exceeds a fixed threshold.  
For consistency with previous work on counterparty loss \cite{Gai10,Gleeson11,Gleeson11b}, we set this threshold to $5\%$.  The contagion probability is then measured as the fraction of runs in which a global cascade results from the initial shock.  The conditional average extent of contagion is the fraction of failed banks, averaged \emph{only} over those runs where a global cascade occurs.

\subsection{Effect of diversification and crowding}

\begin{figure}[h]
\begin{center}$
\begin{array}{cc}
\includegraphics[width=2.5in]{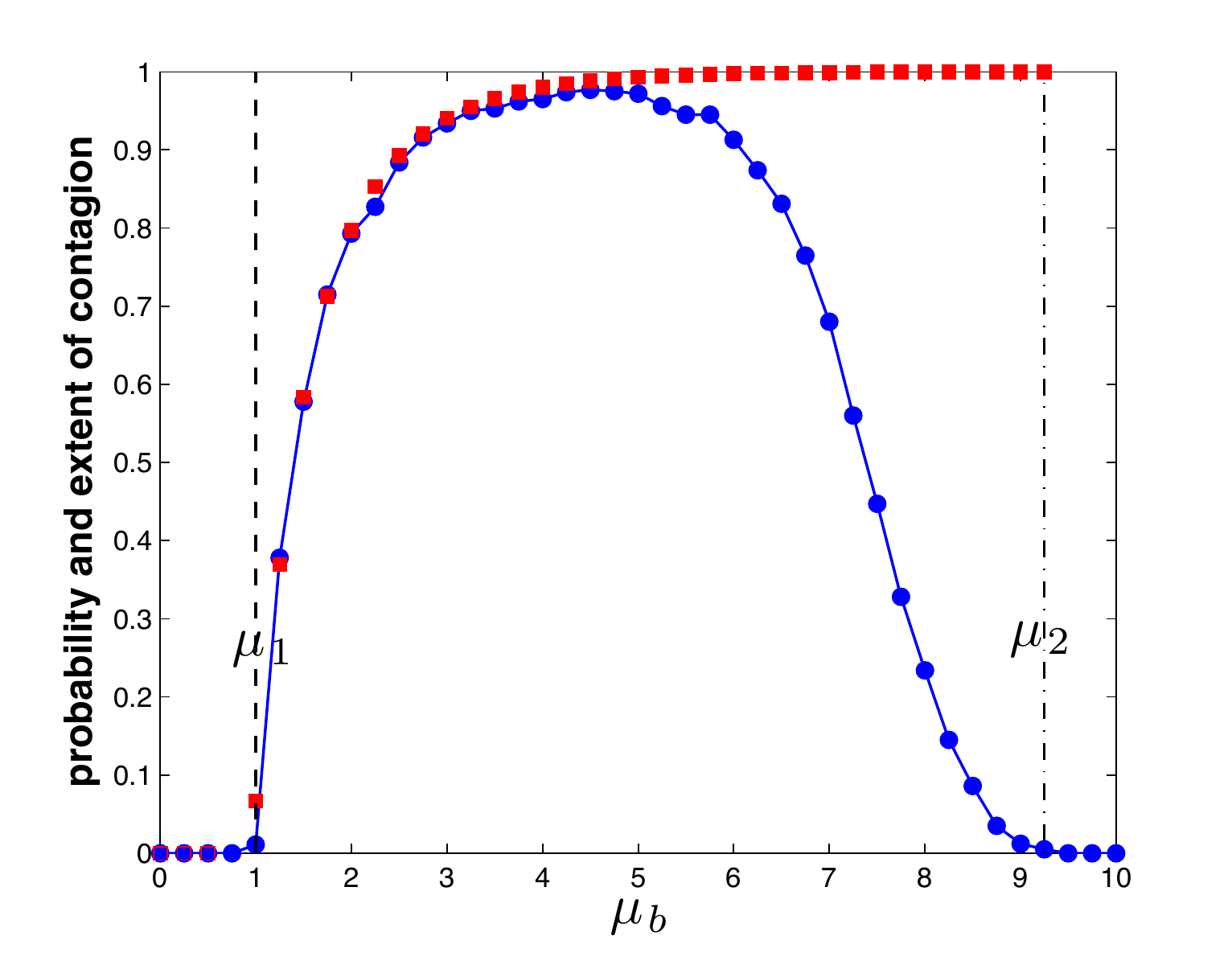}
\includegraphics[width=2.5in]{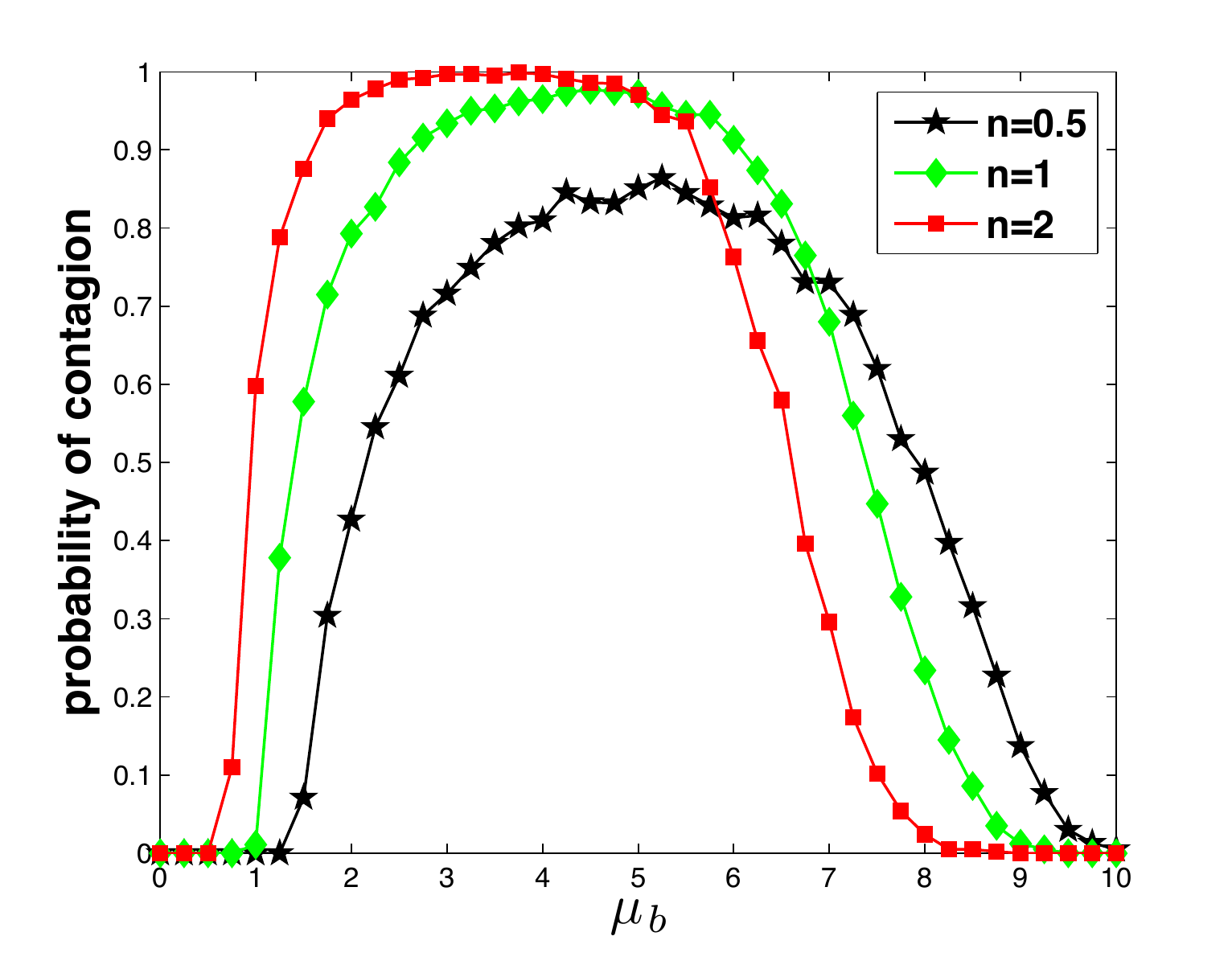}
\end{array}$
\end{center}
\caption{\footnotesize{{\bf Left panel:} contagion probability (blue dots, the solid line is a guide for the eye) and conditional extent of contagion (red squares) measured from $1000$ simulations of a system with $N=M=10^4$. 
In each run, the initial shock consists of dropping the price of a random asset by $35\%$ at the beginning of the simulation.  
We vary the average degree of diversification $\mu_b = \mu_a$.  The two vertical dashed lines mark our numerical estimates for the critical values $\mu_1$ and $\mu_2$ where phase transitions occur, and show the existence of a contagion window between these transitions where global cascades occur with non-zero probability. The system also displays a ``robust yet fragile'' behavior for $\mu_b$ slightly below $\mu_2$: the probability of a global cascade is small, but when one occurs it affects almost all the banks.  {\bf Right panel}: contagion probability for systems with $N=10^4$ and $M=5\times10^3$ (red squares), $M=10^4$ (green diamonds) and $M=2\times 10^4$ (black stars) as a function of the average banks' degree. Solid lines are a guide for the eye. The boundaries $\mu_1, \mu_2$ of the contagion window depend on the value of the crowding parameter $n=N/M$: for larger $n$ both phase transitions are shifted to the left.}}
\label{fig1}
\end{figure}
We begin with the case where the initial shock consists of devaluing a random asset., and examine the dependence on diversification and crowding.
In the left panel of Figure~\ref{fig1}, we plot the probability and conditional extent of contagion measured for a system of $N=10^4$ banks and $M=10^4$ assets  as a function of the average banks' degree $\mu_b$.   
Results refer to $1000$ runs in which a random asset is initially devalued by $35\%$.  
We observe phase transitions at two critical values $\mu_1, \mu_2$ of $\mu_b$, with a \emph{contagion window} in between where global cascades of failures occur with non-zero probability.  Above and below this window, where $\mu_b > \mu_2$ or $\mu_b < \mu_1$, global cascades do not occur.

The existence of a contagion window, and the nonmonotonicity of the contagion probability as a function of $\mu_b$, can be understood with the following arguments.   On one hand, for sufficiently low values of $\mu_b$, stress cannot propagate through the system because the network is poorly connected; there is not enough overlap between the banks' portfolios to spread the cascade.  In particular, for small enough $\mu_b$ the network of banks and assets consists of small components disconnected from one another, so even if every bank is extremely vulnerable to collapse, an initial shock will only affect one of these components.  Thus there is a critical $\mu_1$ below which the cascade cannot propagate; the initial shock might affect a few nearby banks, but the cascade quickly dies out.

On the other hand, if the banks' portfolios are sufficiently diverse, they are robust with respect to devaluing any single asset in their portfolio.  Moreover, a larger average bank degree $\mu_b$ also implies a larger average asset degree $\mu_a$, so each institution typically holds a smaller fraction of the shares of any given asset.  As a consequence, each bank failure has a relatively small effect on asset prices, and most banks remain solvent even if some of their assets are devalued.  Thus there is a critical $\mu_2$ above which cascades quickly die out even though the network is highly connected.

The left panel of Figure~\ref{fig1} also shows that the system displays a ``robust yet fragile'' behavior \cite{Gai10} for some values of the parameters.  Specifically, if $\mu_b$ is slightly less than $\mu_2$, just inside the upper end of the contagion window, the probability of a global cascade is very small, tending continuously to zero as $\mu_b$ approaches $\mu_2$ from below.  But when a global cascade does occur, it affects almost all the banks: the conditional extent of the contagion is almost $1$.  

In the right panel of Figure~\ref{fig1} we plot the contagion probability for different values of the crowding parameter $n=N/M$.  As $n$ increases, the contagion window shifts to the left, decreasing both $\mu_1$ and $\mu_2$.  

The shift in $\mu_1$ can be understood in terms of the appearance of a giant connected component in the network.  In the ensemble of random networks considered here, the emergence of the giant component corresponds to the situation where the average number of banks to which a given bank $b$ is exposed, i.e.\ the average number of banks whose portfolios share at least one asset with $b$, is one. Equivalently, this is the average degree of the \emph{projected network} where two banks are connected if they share an asset.  For this ensemble (essentially the bipartite version of the Erd\H{o}s-R\'enyi model) this degree is $\mu_a \mu_b = \mu_b^2 n$, giving $\mu_1 = 1/\sqrt{n}$.  

To explain the shift in the second transition point $\mu_2$, we note that the drop in price of an asset caused by the liquidation of a portfolio is a decreasing function of $n$. This is because the average number of banks investing in a given asset is $\mu_a=\mu_b n$.  If each bank owns a smaller fraction of an asset, the market impact of a fire sale on that asset is smaller.  When $n$ is larger, this effect takes over at a smaller value of $\mu_b$.

Note that, as a result, changing the crowding parameter $n$ has different effects on the network's stability depending on the value of $\mu_b$.  If $\mu_b$ is close to $\mu_1$, increasing $n$ while keeping $\mu_b$ fixed increases the instability of the system, moving it into the contagion window by increasing the connectivity of the network.  The opposite is true if $\mu_b$ is close to $\mu_2$, where increasing $n$ moves us outside the contagion window by making assets and banks more robust.  Thus the contagion probability is not a monotonic function of $n$.

\subsection{Dependence on shocks}

\begin{figure}[h]
\begin{center}
\includegraphics[width=4.0in]{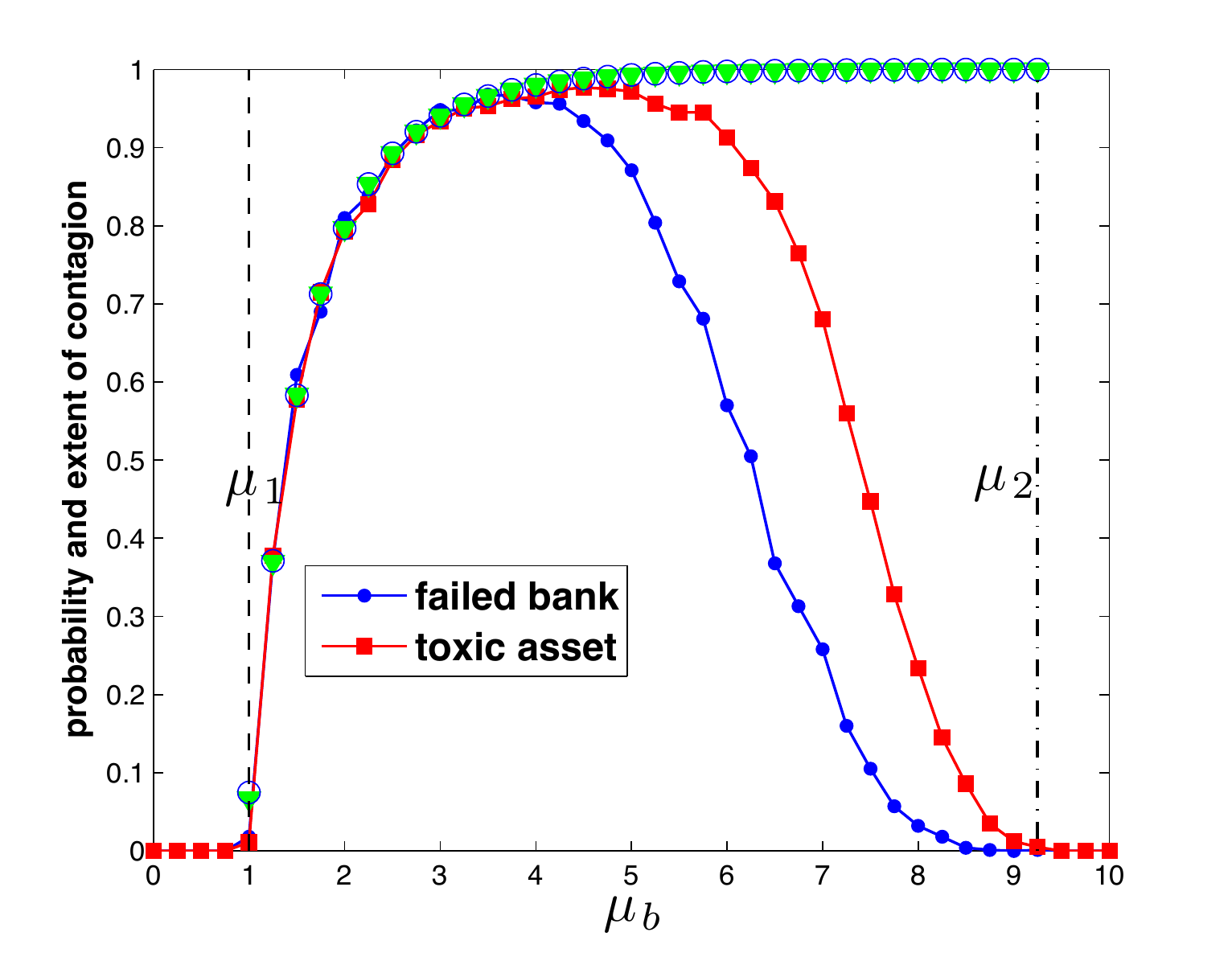}
\end{center}
\caption{\footnotesize{The probability of contagion, and the average conditional extent of contagion, as a function of $\mu_b$ for the two types of initial shock (failed asset vs.\ failed bank).  
Red squares: contagion probability where a random asset is devalued by $35\%$.  
Blue dots: the contagion probability when a random bank fails.  
Blue circles and green triangles: conditional extent of contagion for asset shocks and bank shocks respectively. We see that while the probability of contagion differs between the two types of shocks, the window $\mu_1 < \mu_b < \mu_2$ in which they occur with non-zero probability is the same.  Moreover, when a global cascade does occur, its average size is the same for both types of shocks. 
Results refer to $1000$ simulations of systems with $N=M=10^4$.}}
\label{fig5}
\end{figure}

The above simulations started with an initial shock consisting of devaluing a random asset.  
We now consider the case where we begin with the failure of a random bank.  Figure \ref{fig5} shows a comparison between simulations with shocks of these two types.  
We observe that the probability of contagion depends on the type of shock, but the contagion window and the conditional extent of contagion are the same for both types of shock.  The reason is simple: while the initial conditions of these two processes are different, their dynamics are the same.  Once a cascade has begun, it doesn't matter what kind of shock began it.  Thus the region where the dynamics cause a cascade to spread rather than die out is the same in both cases, as is the eventual size of a global cascade if one occurs.

\begin{figure}[h]
\begin{center}$
\begin{array}{cc}
\includegraphics[width=2.5in]{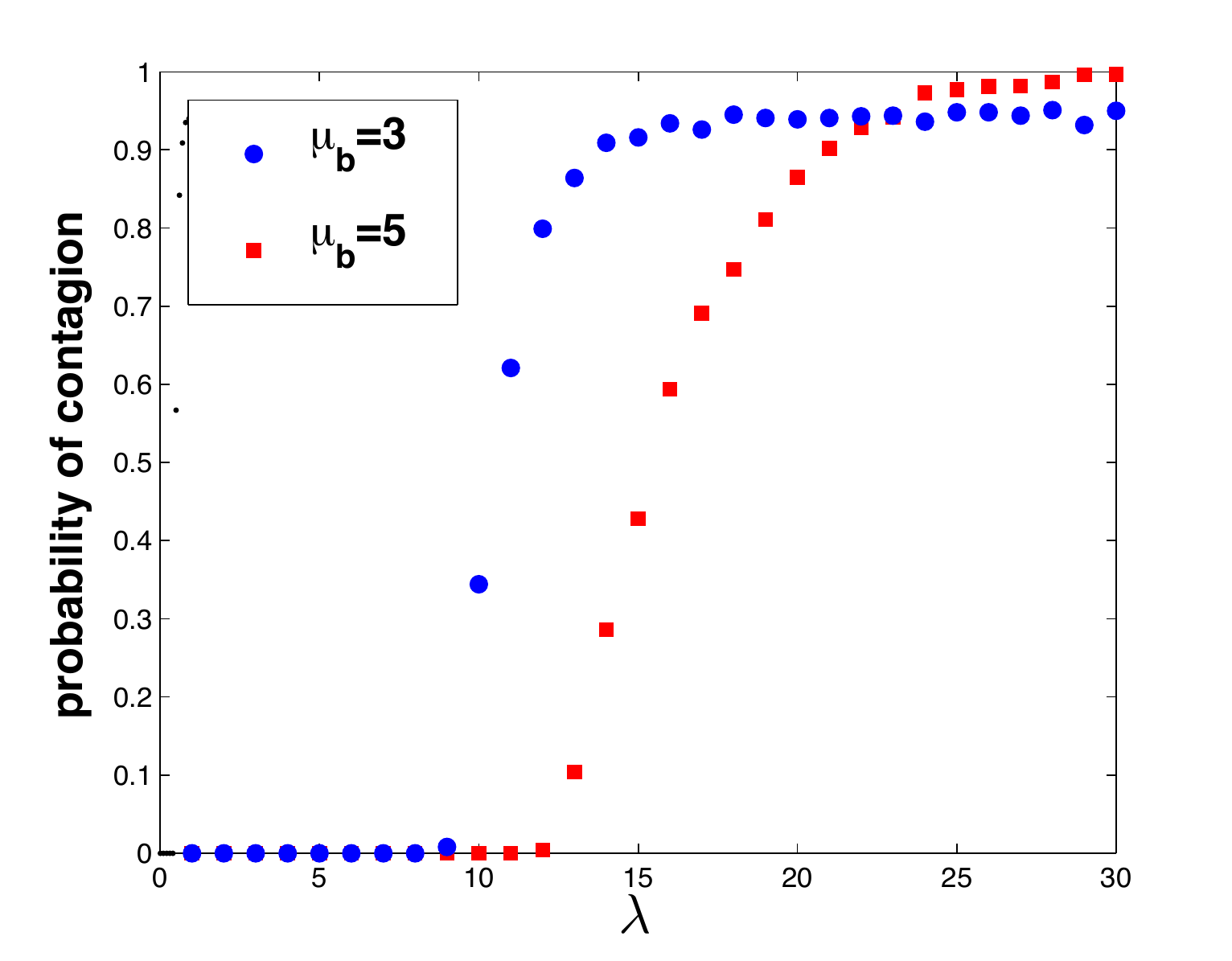}
\includegraphics[width=2.5in]{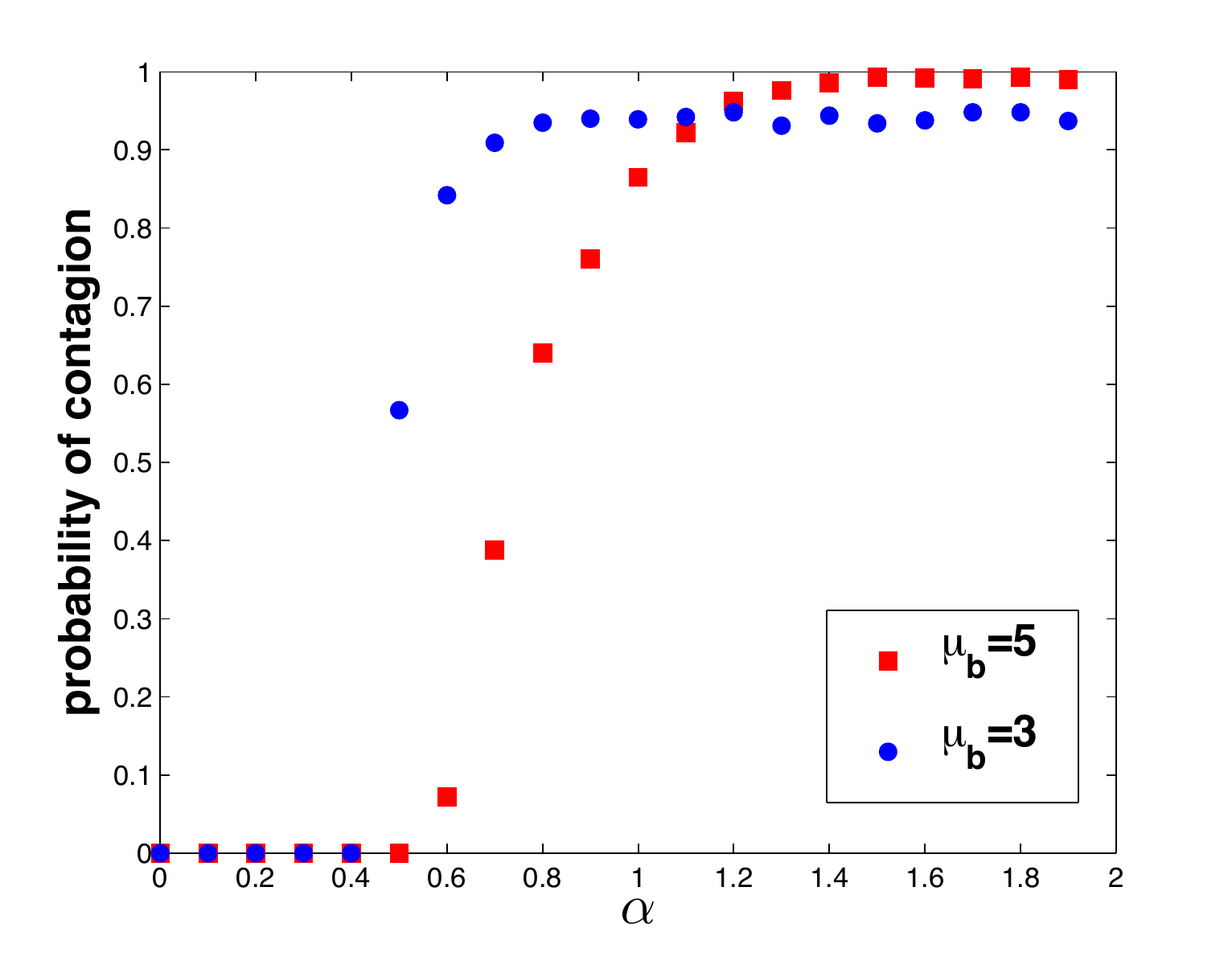}
\end{array}$
\end{center}
\caption{\footnotesize{{\bf Left panel:} Contagion probability as a function of leverage measured from $1000$ simulations of a system with $N=10000$ for different values of $\mu_b$.The initial shock considered is the initial failure of a random bank. Contagion probability is a monotonic function of leverage, and a phase transition separates a regime where no global cascades are observed from one where they occur with non-zero probability. {\bf Right panel:} Contagion probability as a function of the market impact parameter $\alpha$. Increasing market impact has a similar effect as increasing leverage.
}}
\label{lev}
\end{figure}

\subsection{Leverage}  We now show what happens for different values of initial leverage. In the left panel of Figure~\ref{lev} we plot the contagion probability for different values of $\mu_b$ as a function of $\lambda$.  We observe that, for each $\mu_b$, there is a critical value of $\lambda$ above which global cascades occur with non-zero probability, and below which they do not.  This is of interest for regulatory purposes, since it implies the existence of a critical level of leverage below which systemic stability is guaranteed.  In addition, the critical value of $\lambda$ increases as $\mu_b$ increases: in other words, increasing diversification allows for a greater degree of leverage without creating systemic events. 

In the right panel of Figure~\ref{lev}, we show that a similar behavior occurs as we change the parameter $\alpha$ that appears in the market impact function while keeping the leverage fixed.  That is, for a given value of $\mu_b$ and $\lambda$, there is a critical value of $\alpha$ above which contagion occurs.  This is not unexpected, since under the assumptions specified in Section~\ref{assumptions} the solvency condition for bank $i$ can be written as 
\be
\lambda_i\le\frac{\sum_{j=1}^M Q_{ij} \e^{-\alpha x_j^t}}{E_i^0}+1,
\ee
where $x_j^t$ is the fraction of shares of asset $j$ liquidated up to time $t$.  When $\alpha$ is larger, the market impact of a fire sale is greater, causing a sharper drop in asset prices.  On the other hand, increasing diversification $\mu_b$ increases this critical value of $\alpha$,  showing that diversification allows banks to survive a larger price impact.

Summarizing, we presented in this section results of numerical simulations for bipartite networks with Poisson degree distributions for both banks and assets.  The probability and the average extent of contagion have been measured for two different types of shocks, namely the initial depreciation of a random asset or the initial failure of a random bank. Our simulations suggest that:
\begin{itemize}
\item As a function of the average diversification of banks' portfolios, represented by their average degree $\mu_b$, the system is characterized by two phase transitions that define a contagion window where global cascades occurs with non-zero probability.
\item Changing the crowding parameter $n$, i.e.\ the ratio of the number of banks to the number of assets available for investment, can increase or decrease the contagion probability depending on which of these transitions we are close to. 
\item Although the contagion probability is different for the two types of initial shocks, the contagion window within which global cascades occur, and the average extent of these cascades when they occur, are the same.
\item The system displays a ``robust yet fragile'' behavior, with regions in parameter space where global cascades are very unlikely, but where almost the entire system is affected if one occurs.
\item For each fixed $\mu_b$ and $n$, there is a critical value of the leverage $\lambda$ above which the system becomes unstable.  This critical value of $\lambda$ increases with $\mu_b$.
\end{itemize}

\section{Comparison to predictions from stability analysis}

\begin{center}
\begin{figure}[h]
\begin{center}
\includegraphics[width=4in]{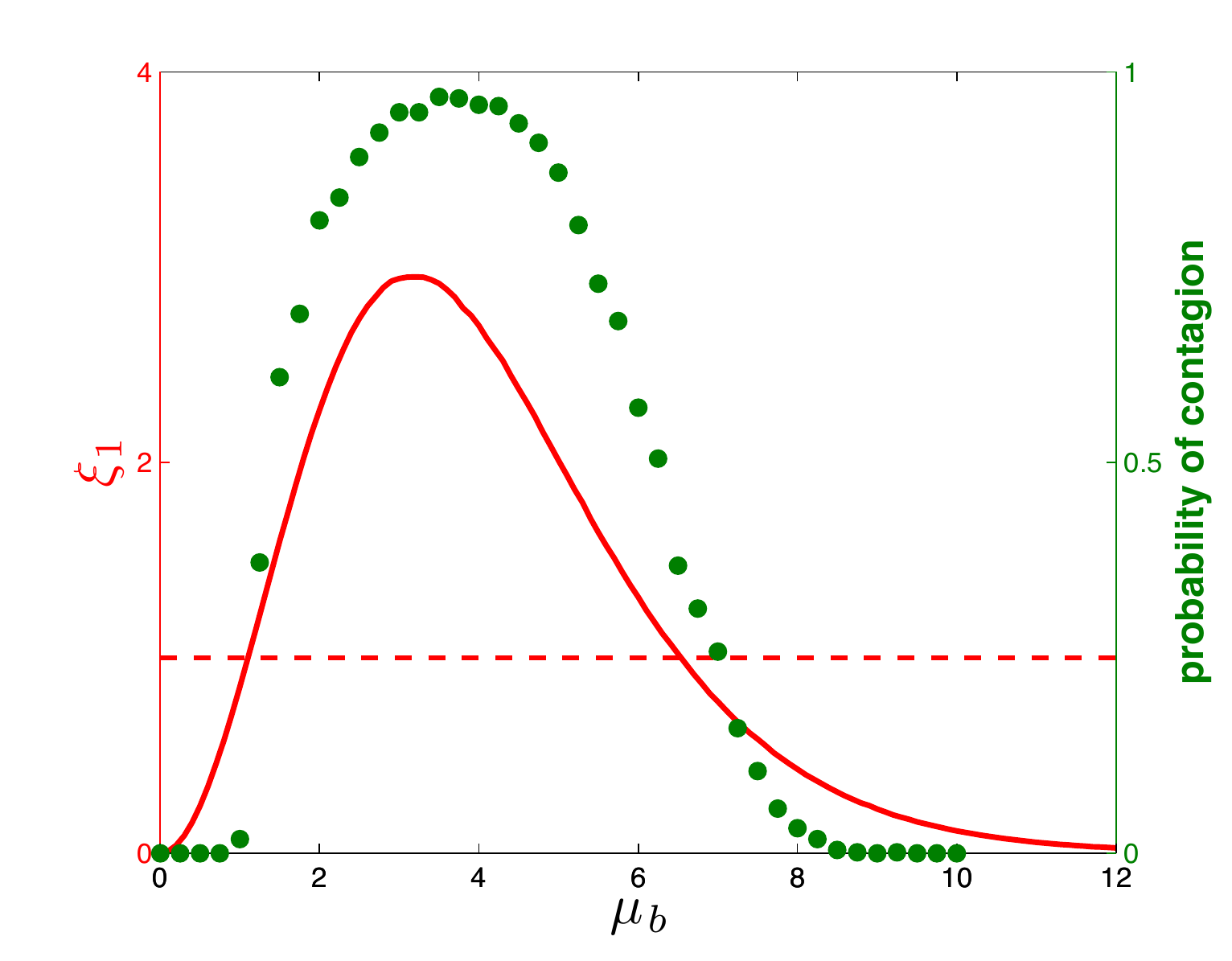}
\caption{\footnotesize {\textit{ { Contagion probability (green dots, right axis) as computed from numerical simulations of a system of size $N=M=10^4$. The red solid line (left axis) represent the largest eigenvalue $\xi_1$ of the matrix $\mathcal{N}$. The dashed horizontal line is in correspondence to $\xi_1=1$. If $\xi_1>1$ global cascades are observed in numerical simulations. The theory underestimates the width of the contagion window, as it only gives a sufficient condition for global cascades to occur. However, the discrepancy between theory and numerical results is partly due to finite size effects (see Figure~\ref{theoryfig2}).}}}}
\label{theoryfig1}
\end{center}
\end{figure}
\end{center}

\begin{center}
\begin{figure}[h]
\begin{center}
\includegraphics[width=8cm]{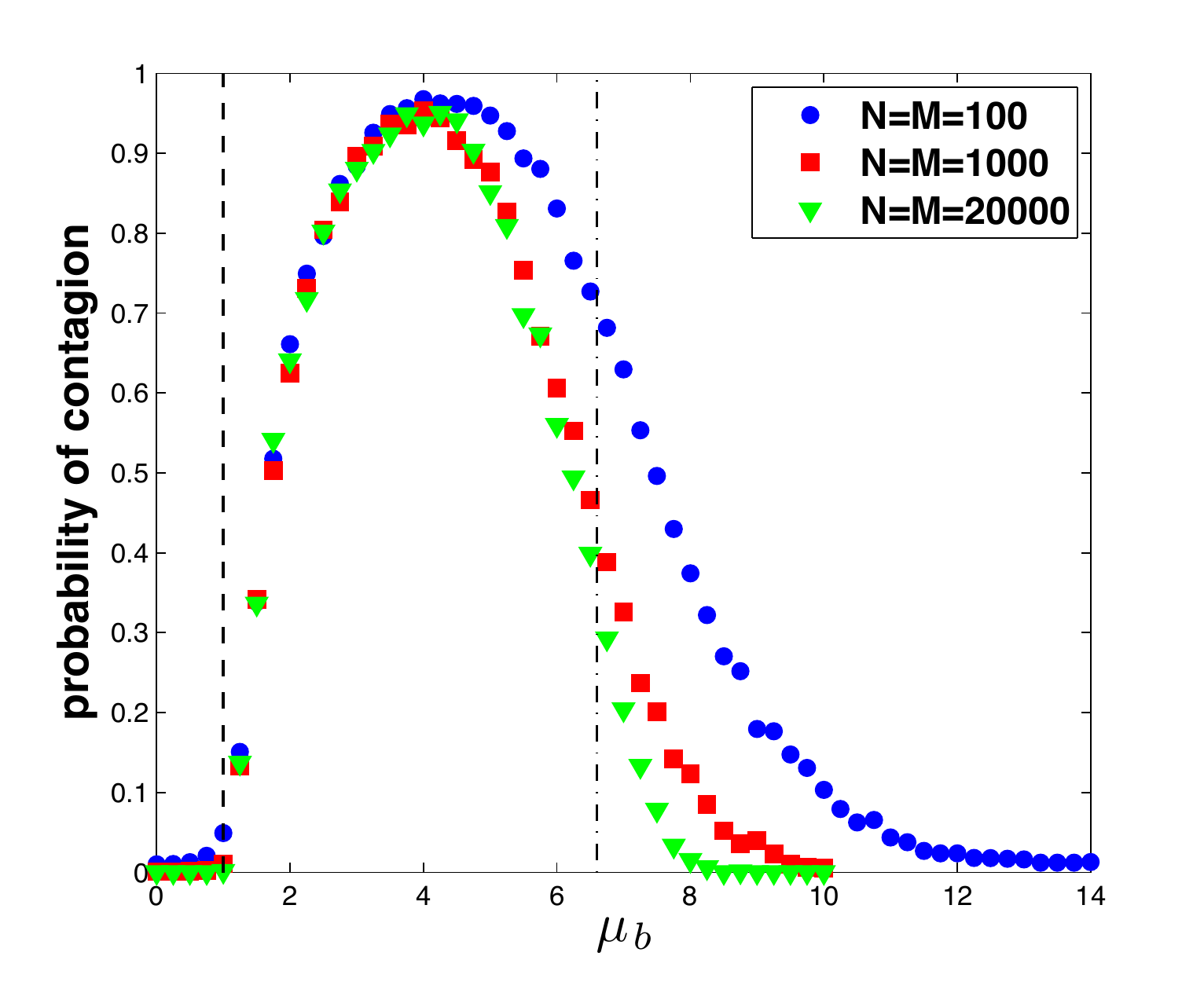}
\caption{\footnotesize {\textit{ { Simulation results for $N=100$ ( blue circles),  $N=1000$ (red squares), $N=20000$ (green triangles), $n=1$.  The vertical dashed lines are drawn in correspondence to the phase transitions predicted by the analytic calculation. As the size of the system increases the agreement between theory and simulations improves. Finite size effects are expected given that the theory is valid in the limit $\{N,M\}\to\infty$}. 
}}}
\label{theoryfig2}
\end{center}
\end{figure}
\end{center}

We now compare the numerical results presented in the previous section to those based on stability analysis.  The stability analysis depends on two assumptions that are not necessarily well-satisfied in the simulation.  The first is that $M$ and $N$ are both infinite (even though their ratio $n = N/M$ is finite), and the second is that the failure process can be described through a branching process on a tree.

We estimated $F(h,k,\ell)$ through monte-carlo methods and assumed that the contribution coming from banks with degree higher than $200$ is negligible.   We then numerically diagonalized the $200$ by $200$ matrix $\mathcal{N}$. We discuss in the following the results obtained in the case where $P_a(\ell)$ and $P_b(h)$ are Poisson distributions.

In Figure~\ref{theoryfig1} we plot for $n=1$ the largest eigenvalue of $\mathcal{N}$ and we compare it with the contagion probability as computed from numerical simulations. As expected, when the largest eigenvalue of $\mathcal{N}$ is greater than $1$ global cascades are observed.  We see from the figure that the analytic calculation underestimates the size of the contagion window. This is partly due to finite size effects, as observed for instance in \cite{Watts02}. We plot in Figure~\ref{theoryfig2} the contagion probability as measured from numerical simulations for $n=1$ and different values of $N$. From the figure we clearly see that by increasing the size of the system the discrepancy between analytic and numerical calculations gets smaller, and that the analytic solution, although giving only a sufficient condition for global cascades to occur, produces a reasonable estimate of the contagion window when $N$ and $M$ are large. 

 \begin{figure}[h]
\begin{center}$
\begin{array}{cc}
\includegraphics[width=3.0in]{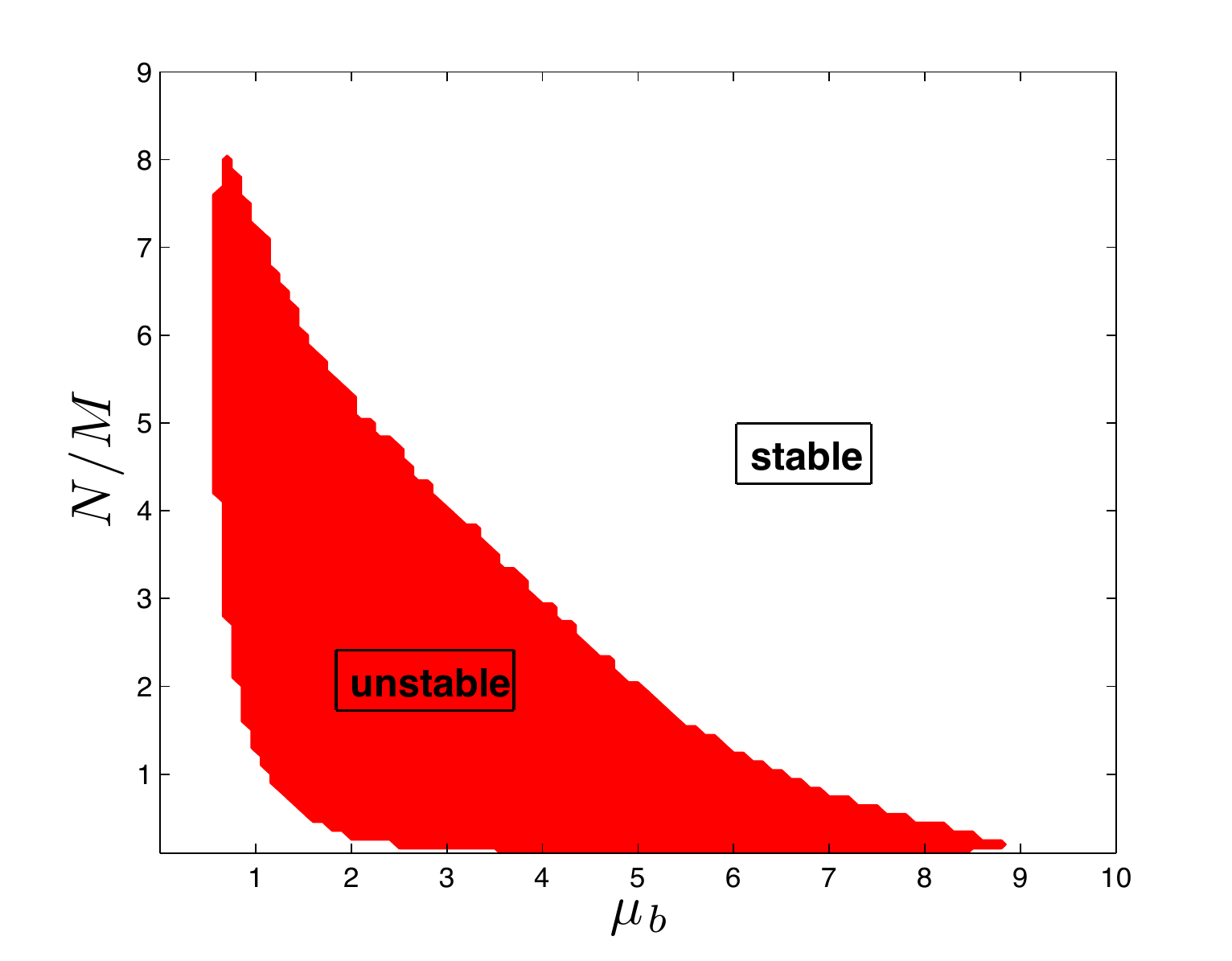}
\includegraphics[width=3.2in]{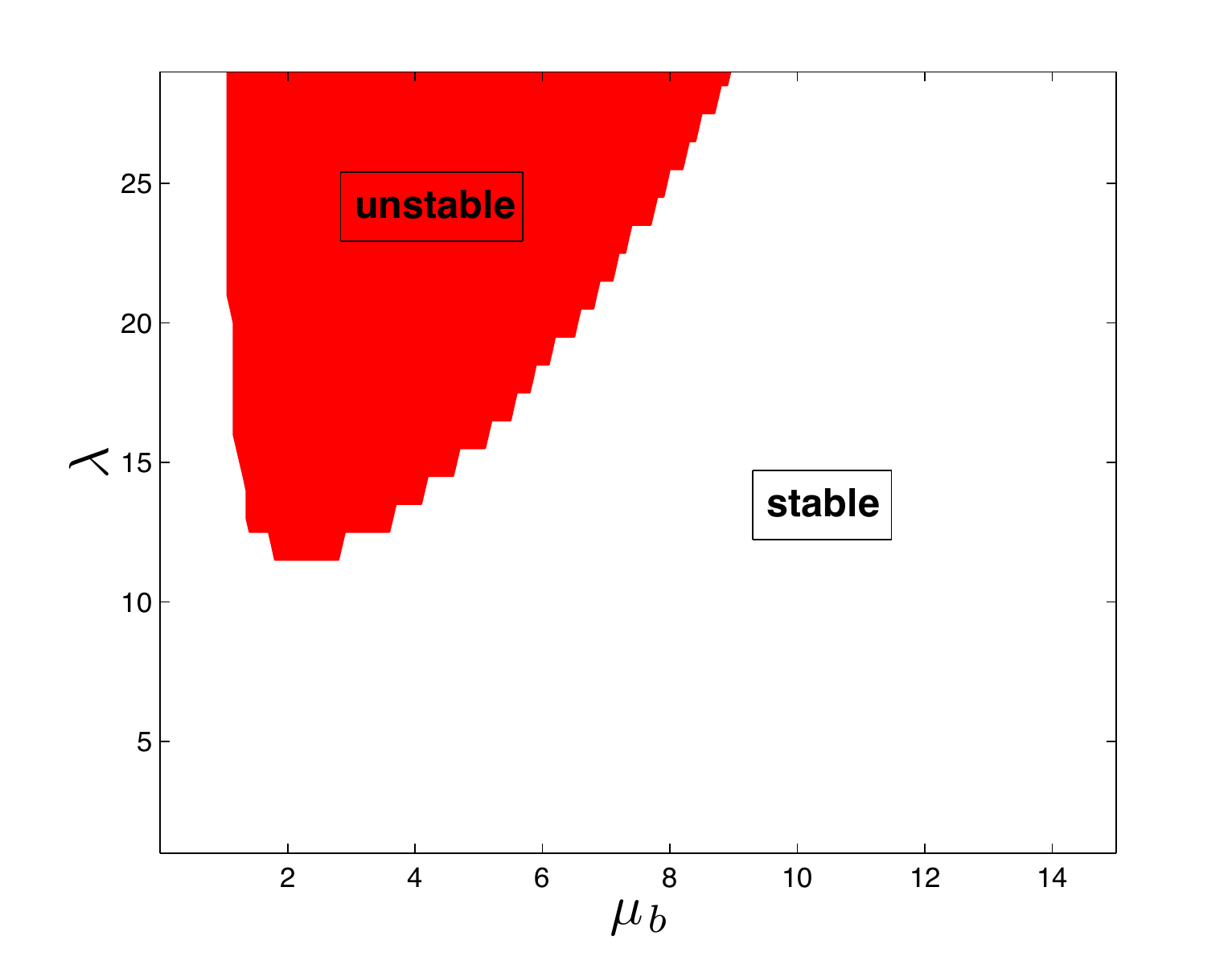}
\end{array}$
\end{center}
\caption{\footnotesize{{\bf Left panel}: The red region is the region of phase space where global cascades occur for a system with $\lambda=20$ as a function of $\mu_b$ and $n$. {\bf Right panel}: the red region is the region of phase space where global cascades occur for a system with $n=1$ as a function of $\mu_b$ and $\lambda$. Points refer to the phase transition as measured from the largest eigenvalue of $\mathcal{N}$}. Lines are a guide for the eye.}
\label{theoryfig3}
\end{figure}

 \begin{figure}[h]
\begin{center}
\includegraphics[width=4.5in]{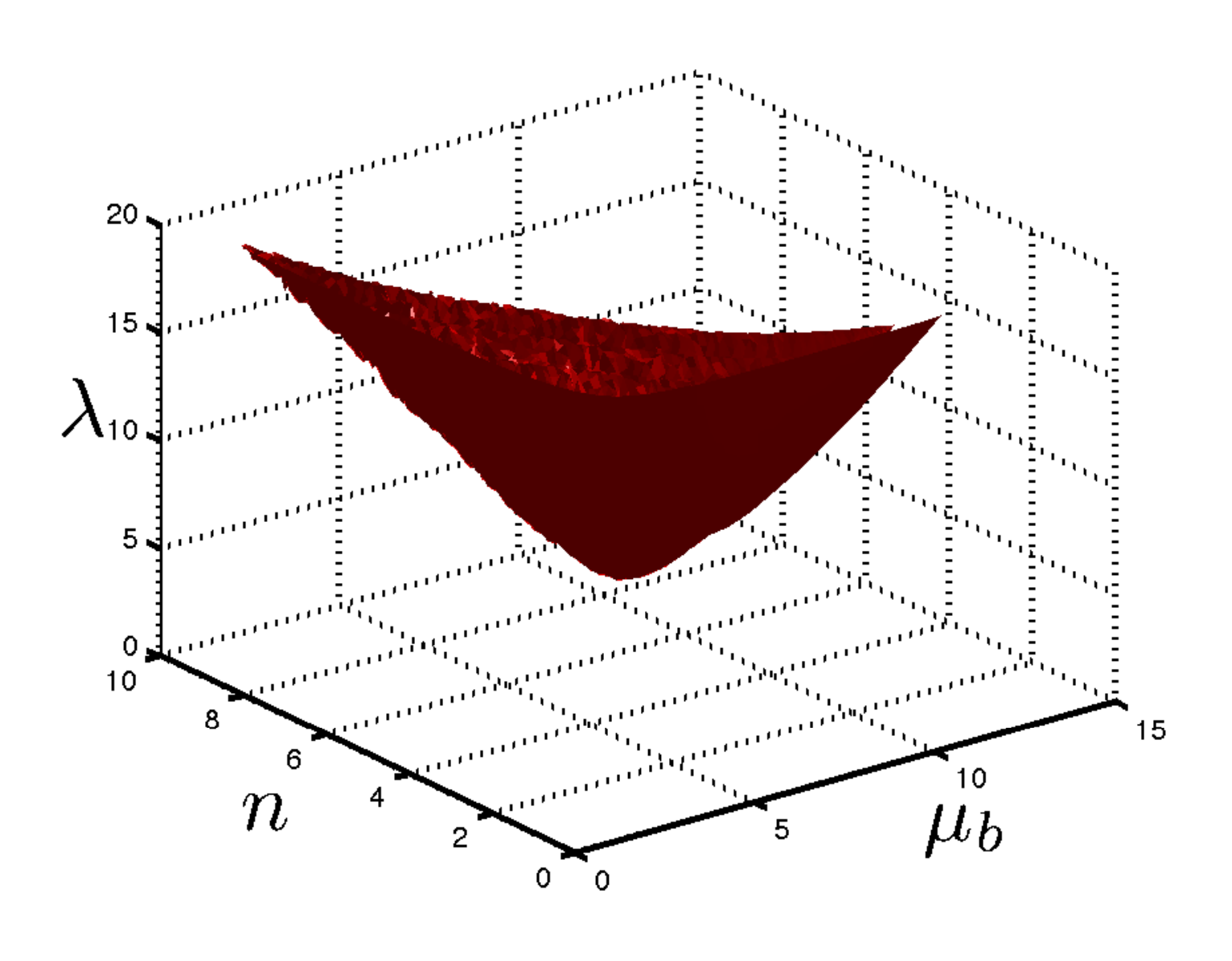}
\end{center}
\caption{\footnotesize{ 3D visualization of the region of parameter space where global cascades occur with non-zero probability as predicted with the analytical approach. Global cascades are observed within the cone-shaped colored region. We note in particular that there is critical value of leverage below which global cascades do not occur for any values of diversification and crowding parameter.}}
\label{phase3d}
\end{figure}

We finally plot in Figure~\ref{theoryfig3} the phase diagram obtained with our analytic approach. The region within the solid line in the left panel represent the region where global cascades occur for $\lambda=20$. From the figure we can see the features already observed in numerical simulations. In particular, for fixed $n$, we clearly see the existence of two phase transitions that define a window of connectivities where global cascades occur with non-zero probability. As we change $n$, the analytic calculation also predicts the shift in the transition points, that tend to move to higher values of $\mu_b$ as $n$ is decreased. In the right panel of Figure~\ref{theoryfig3}, we depict the phase diagram for $n=1$ as a function of leverage $\lambda$ and average diversification $\mu_b$. As expected, we see that the contagion window widens as $\lambda$ increases. A three dimensional visualization of the phase diagram is reported in Figure~\ref{phase3d}, where the red region denotes the unstable region of parameter space. 
Interestingly, we observe the existence of a minimum level of leverage ($\lambda\simeq12$) that leads to the occurrence of global cascades of failure. This feature is of potential interest for regulators, since it is equivalent to the existence of a maximum level of leverage below which the system is overall stable.

Another point of potential interest for regulators is the eigenvalue of the matrix $\mathcal{N}$. In a dynamic setting in which banks operate under no stress circumstances, one expects $\xi_1$ to change over time as banks trade to rebalance their portfolios. By monitoring the time behavior of $\xi_1$, a regulator would notice if the system is approaching a dangerous regime as $\xi_1$ gets closer to $1$, and could act to increase the stability of the system.

\section{Conclusion}

We have developed a framework for thinking about the stability properties of banking networks due to overlapping portfolios.  This framework emphasizes that the key property is stability:  If the system is stable, shocks will not propagate; if it is unstable, a shock can be amplified and trigger cascading bankruptcies.  This can be discussed in terms of a branching process that gives insight into the dynamics of failure.  While we have called these ``banking networks" for simplicity, the basic ideas are relevant for any leveraged financial institutions.

To understand how the stability of banking networks might depend on parameters such as diversification, leverage and crowding, we formulated a stylized model of a financial system in which $N$ banks with average diversification $\mu_b$ invest in a common pool of $M$ assets. The system can be conveniently described in terms of a bipartite network, with banks being connected through links to the assets in their portfolios. Links have a two-fold role in such a network. On one hand, they allow individual banks to diversify their investment and reduce their exposure to a specific asset. On the other hand, they are channels for the propagation of financial contagion.  We characterized the response of such system to initial shocks affecting a single asset or bank.
 
The relevant parameters for the model are the average diversification $\mu_b$, the crowding parameter $n=N/M$ (that measures the proportion of banks to assets), and the initial leverage $\lambda$. By means of numerical simulations we showed the existence of phase transitions separating a region in parameter space where global cascades occur from a region where global cascades never occur. In particular, the double role played by links in the bipartite network representing the system is reflected in a nonmonotonic behavior of the contagion probability as a function of $\mu_b$, where we observe the existence of two phase transitions at $\mu_b=\mu_1$ and $\mu_b=\mu_2$ that define a window of connectivities such that global cascades occur if $\mu_1\le\mu_b\le\mu_2$. Changing the crowding parameter $n$ has the effect of shifting the location of the phase transitions. Finally, our model shows that increasing leverage
increases the overall instability of the system, but that there is a critical level of leverage below which global cascades do not occur for any value of diversification or crowding.

Using an analytical approach based on generalized branching processes on networks, we are able to analytically estimate the region of parameter space where global cascades occur.  This branching process is different from standard ones in the fact that the fate of a node depends on its degree and on the degree of all its neighbors. This greatly increases the difficulty of the problem.  We are nonetheless able to solve it by generalizing existing methods.  Thus, apart from their specific application to financial contagion, our methods can be applied to a wide variety of contagion models, where susceptibility and transmission probabilities depend on node degrees. 

The mechanistic model considered in this paper can be extended in several directions. First of all, it would be interesting to relax some of the specific assumptions considered in this paper (homogeneity of banks' balance sheets, Poisson degree distributions, market impact function) in order to understand how different choices for the network topology or the statistical properties of balance sheets impact the stability of the system. Although we do not expect different results from a qualitative point of view, it should nonetheless be possible to asses the relative stability of systems with different properties, similarly to what has been done for counterparty loss in \cite{Caccioli12}. In particular, it would be very useful to empirically characterize real systems and calibrate the model with real data. This could potentially make it possible to test the effectiveness of new policies aimed at reducing systemic risk.

A further direction we plan to pursue in the future is to go beyond the mechanistic model by considering a more realistic price dynamics and allowing banks to react to price fluctuations by rebalancing their portfolios. This should allow the system to develop endogenous crisis similar to the ones observed in  \cite{Thurner12}, and to generate the systemic instabilities induced by leverage and mark-to-market accounting practices discussed in \cite{Caccioli12b}.

\section*{Acknowledgments}
This work was supported by the National Science Foundation under grant 0965673,
by the European Union Seventh Framework Programme FP7/2007-2013 under
grant agreement CRISIS-ICT-2011-288501 and by the Sloan Foundation. C.M. is supported by the AFOSR and DARPA under grant \#FA9550-12-1-0432. The authors would like to thank Adam Ashcraf, Morten Bech and Martin Summer for useful discussions.

\bibliography{ref}{}
\bibliographystyle{ieeetr}

\end{document}